\documentclass{article}
\usepackage{ijcai23}
\usepackage{tikz}
\usepackage{times}
\usepackage{soul}
\usepackage{url}
\usepackage{todonotes}
\usepackage[hidelinks]{hyperref}
\usepackage[utf8]{inputenc}
\usepackage[small]{caption}
\usepackage{graphicx}
\usepackage{amsmath}
\usepackage{amsthm}
\usepackage{booktabs}
\usepackage{algorithm}
\usepackage{algorithmic}
\usepackage{glossaries}
\usepackage[switch]{lineno}
\usepackage{svg}
\usepackage{booktabs}
\usepackage{graphicx}
\usepackage{svg}
\usepackage{makecell}
\usepackage{adjustbox}
\usepackage{comment}
\newacronym{ai}{AI}{Artificial Intelligence}
\newacronym{cpra}{CPRA}{California Privacy Rights Act}
\newacronym{dsp}{DSP}{Digital Signal Processing}
\newacronym{fl}{FL}{Federated Learning}
\newacronym{gdpr}{GDPR}{General Data Protection Regulation}
\newacronym{mos}{MOS}{Mean Opinion Score}
\newacronym{nonIID}{non-IID}{non-independent and identically distributed data}
\newacronym{xai}{XAI}{Explainable Artificial Intelligence}
\newacronym{mfcc}{MFCC}{Mel Frequency Cepstral Coefficients}
\newacronym{tts}{TTS}{Text-To-Speech}
\newacronym{vc}{VC}{Voice Conversion}
\newacronym{cnns}{CNNs}{Convolutional Neural Networks}
\newacronym{stft}{STFT}{Short-time Fourier Transform}
\newacronym{dnn}{DNN}{Deep Neural Network}
\newacronym{la}{LA}{Logical Access}

\title{Uncovering the Deceptions: An Analysis on Audio Spoofing Detection and Future Prospects}

\author{
Rishabh Ranjan
\and
Mayank Vatsa\and
Richa Singh
\affiliations
Indian Institute of Technology, Jodhpur, India\\
\emails
\{ranjan.4, mvatsa, richa\}@iitj.ac.in
}

\begin{document}

\maketitle

\begin{abstract}
Audio has become an increasingly crucial biometric modality due to its ability to provide an intuitive way for humans to interact with machines. It is currently being used for a range of applications including person authentication to banking to virtual assistants. Research has shown that these systems are also susceptible to spoofing and attacks. Therefore, protecting audio processing systems against fraudulent activities such as identity theft, financial fraud, and spreading misinformation, is of paramount importance. This paper reviews the current state-of-the-art techniques for detecting audio spoofing and discusses the current challenges along with open research problems. The paper further highlights the importance of considering the ethical and privacy implications of audio spoofing detection systems. Lastly, the work aims to accentuate the need for building more robust and generalizable methods, the integration of automatic speaker verification and countermeasure systems, and better evaluation protocols. 
\end{abstract}

\section{Introduction}
Voice as a  biometric modality has been used for both identification and verification tasks, and it has a wide range of real-world applications such as banking, government, and law enforcement operations. For instance, Citi\footnote{\url{https://tinyurl.com/citibankvoice}} uses voice samples to authenticate persons using automatic speaker verification (ASV) systems. Voice-enabled devices and virtual assistants such as Google Home and Microsoft Cortana are commonly used at home to help us manage our daily tasks. The worldwide voice recognition market has surpassed $\$$3.5 billion in 2021 and is expected to reach $\$$10 billion by 2028\footnote{\url{https://tinyurl.com/voicemarket}}. With the host of applications, there come several challenges associated with it. These voice-enabled devices often store a large amount of personal information and speech samples, and this data can be used to imitate one's voice. In 2020, an employee cloned the voice of the company's CEO and committed a financial fraud of $\$$35 million. 
With the advancements in deep learning, it has become very easy to clone someone's voice (e.g., it has been claimed that Microsoft's VALL-E\footnote{\url{https://tinyurl.com/vallemicro}} can clone someone's voice in 3 seconds). 

\begin{figure}[t!]
    \centering
    \includegraphics[width = 0.45\textwidth]{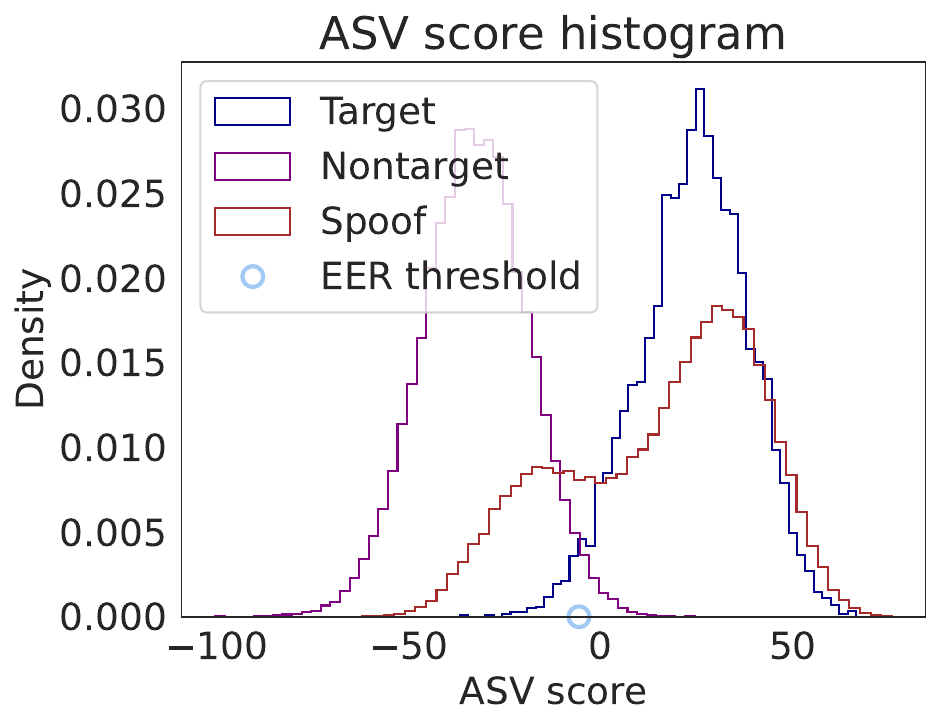}
    \caption{The score distribution of target users, non-target users and spoofed imposters. Due to an overlap in the distribution between the target and the spoofed imposter, models misclassify between bonafide and spoofed samples. The scores are taken from the ASVSpoof2019 dataset.} 
    \label{fig:distribution_score}
\end{figure}

ASV systems \cite{DBLP:journals/speech/KinnunenL10} authenticate a speaker's identity by analysing their voice's unique characteristics. These systems work by analysing the characteristics of a person's voice, such as pitch, tone, and accent, to determine if they are who they claim to be. One of the main challenges with using voice as an identity verification method is the possibility of spoofing attacks \cite{DBLP:conf/interspeech/EvansKY13,DBLP:series/acvpr/EvansKYWAL14,ranjan_ijcb}. The speaker verification scores from the ASVspoof2019 dataset, as shown in Figure \ref{fig:distribution_score}, illustrate that the score distribution of the spoofed imposter is similar to the target user. This suggests that spoofed samples can be easily authenticated as the target users.   

\begin{figure}[t!]
\centering
  \includegraphics[width= 0.485\textwidth]{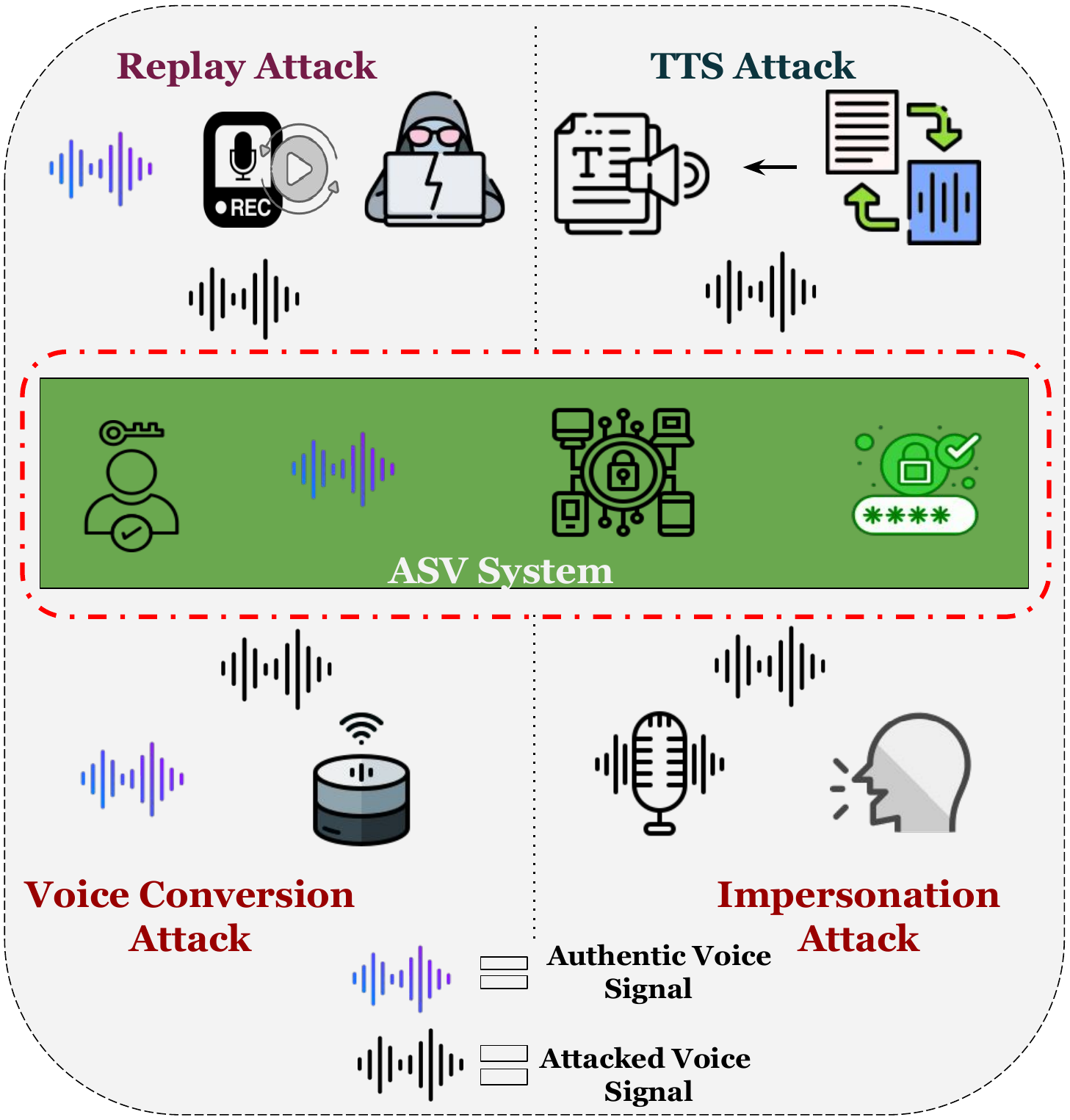}
  \caption{Illustration of different kinds of attacks on the ASV System. Replay and Impersonation attacks can be considered Physical Access attacks, while TTS and VC attacks are considered Logical Access attacks.}
  \label{fig:attacks}
\end{figure}

Audio spoofing attacks are broadly classified into two categories: physical access attacks and logical access attacks. Physical access attacks are those that are introduced at the microphone (source) level, such as replay attacks and impersonation attacks. The different kinds of attacks are shown in Figure \ref{fig:attacks}. In Replay attacks, an attacker records a legitimate user's voice and then plays it back to the ASV system to gain access. This can be done by recording the user's voice during a legitimate login or by intercepting and recording the user's voice during a phone call. Logical access attacks, on the other hand, are introduced at the transmission level. These types of attacks include text-to-speech attacks and voice-conversion attacks. In text-to-speech attacks, an attacker uses a computer program to generate a voice that mimics a legitimate user's voice. This can be done by providing the text input to the program and generate speech that sounds like the legitimate user's voice. In voice-conversion attacks, the attacker uses a legitimate user's voice to generate an artificial voice to match the characteristics of the targeted speaker's voice. Recent studies \cite{kreuk2018fooling} have shown that ASV systems are also vulnerable to adversarial attacks. Several countermeasures are proposed to detect audio spoofing attacks. Most existing countermeasures are based on detecting artifacts in the generated speech. These artifacts can also be called a model's signature. The countermeasure systems take speech waveform as input and classify the input speech into bonafide or spoof.

To provide a more comprehensive understanding, several survey papers \cite{wu2015spoofing,tan2021survey} are published on audio spoofing detection. These papers have examined different aspects of spoofing attacks and the countermeasures that have been developed to combat them. 
The motivation for this research arises from the ongoing discussions and debates about the most effective methods and techniques for building countermeasure systems. This survey aims to provide a comprehensive overview of the current state-of-the-art in building a countermeasure system and identify the gaps or challenges that still need to be addressed. We summarize the  main contributions as follows: \\
\begin{itemize}
    \item This paper presents an analysis of speech generation and associated detection techniques, 
    \item We provide a summary of evaluation metrics and existing datasets along with their limitations, and
    \item We present a meta-analysis and discuss the open challenges and future prospects in audio spoofing detection. 
\end{itemize}


\section{Evaluation Metrics}
This section briefly introduces several standard performance metrics to understand better how to assess spoofing detection and speaker recognition systems.
\paragraph{Equal error rate (EER): }
EER is a commonly used performance metric for evaluating audio spoofing detection systems. The miss and false alarm rates of the spoofing detection system are increasing and decreasing with respect to different thresholds. The point at which the miss rate and false alarm rate become equal is called an Equal Error Rate (EER).
\paragraph{Detection Error Tradeoff (DET) curve:}
DET curve is a graphical representation of the performance of an audio spoofing detection system. It plots the false acceptance rate (FAR) on the x-axis and the false rejection rate (FRR) on the y-axis. The DET curve is similar to the Receiver Operating Characteristic (ROC) curve, but it is specifically used for detection tasks where the costs of false positives and false negatives are different.

\paragraph{Tandem Decision Cost function:}
The better-performing countermeasure system is not guaranteed to have a more reliable speaker verification performance. However, the CM system's performance directly impacts the ASV system's reliability. Thus, it is required to evaluate the performance of CM and ASV systems jointly. A new metric t-DCF \cite{kinnunen2018t} is proposed, which could bridge the gap between the CM system and the ASV system and can also be used to evaluate the performance of the CM system in isolation from the ASV system. The detection cost function is based upon the costs of missing the target users and falsely accepting the imposter users. The proposed t-DCF metric also considers the spoofing imposters and associated costs with accepting or rejecting them. The CM system and ASV system can be integrated in three ways. The metric considers four kinds of costs: a) the cost of the ASV system rejecting a target trial, b) the cost of the ASV system accepting a nontarget trial, c) the cost of CM rejecting a human trial, d) the cost of CM accepting a spoof trial.
\begin{table*}[t!]
\centering

\resizebox{\textwidth}{!}{%
\begin{tabular}{@{}lcccc@{}}
\toprule
\textbf{Dataset}                          & \textbf{Language}              & \textbf{Attack type}        & \textbf{No of Speakers}                & \textbf{No of Samples}   \\ \midrule
YOHO \cite{kreuk2018fooling}                  & English                        & Mimicry                     & 2                                      & 960                      \\ \midrule
WSJ  \cite{ergunay2015vulnerability}                          & English                        & SS and VC                   & 283                                    & -                        \\ \midrule
SAS   \cite{wu2015sas}                          & English                        & SS and VC                   & Real: 106 and Fake: 106    & 2,12,000                 \\ \midrule
ASVspoof 2015 \cite{wu2015asvspoof}       & English                        & SS and VC                   & Real: 106 and    Fake: 106 & 16651 Real + 255904 Fake \\ \midrule
ASV Spoof Noisy Database                 & English                        & SS and VC                   & Real: 106 and   Fake: 106 & 16651 Real + 255904 Fake \\ \midrule
RedDots \cite{Kinnunen2017RedDotsRA}        & 5  languages          & Replay                      &      89                                  & 3750                      \\ \midrule
ASVspoof 2017 \cite{kinnunen2017asvspoof} & English                        & Replay                      & Real: 42 and Fake: 42       & 3566 Real + 14,466 Fake  \\ \midrule
ASVspoof 2019 \cite{wang2020asvspoof}     & English                        & SS,VC and Replay            & Real: 106 and Fake: 106     & 121971 LA + 22157 PA     \\ \midrule
FOR  \cite{reimao2019dataset}                                     & English                        & SS                          & Real: 140 and Fake: 33      & 195000                   \\ \midrule

ASVspoof2021 \cite{delgado2021asvspoof}   & English                        & SS,VC, Replay and Deepfakes & Real:149 and Fake:149       & -                        \\ \midrule
HAD \cite{yi2021half}                                      & Chinese                        & Human                       & Real: 218 and Fake: 218     & -                        \\ \midrule
WaveFake \cite{frank2021wavefake}                                  & English, Japanese              & TTS                         & Real: 2 and Fake: 2        & 117985                       \\ \bottomrule
\end{tabular}%
}

\caption{Summary of existing audio spoof detection datasets.}
\label{tab:all_dataset}
\end{table*}
\section{Spoofed Speech Generation Techniques}

There are several algorithms to create spoofed speech samples depending on the kind of attack. Recent advancements in deep learning have led to the development of robust and sophisticated models for fake audio generation, such as WaveNet \cite{oord2016wavenet}, and Tacotron \cite{wang2017tacotron}, capable of generating high-quality speech that is difficult to distinguish from real speech. In the following subsections, we discuss the different attack generation technologies.

\subsection{Speech Synthesis} 
Speech synthesis is a technology that generates speech from a given input, such as text or speech. The process of generating speech from text is called text-to-speech synthesis, and changing the characteristics of a speaker's voice to make it sound like another speaker is called voice conversion. A typical speech synthesis model has three stages, the Input Analysis phase, the Acoustic Model phase and the Vocoder phase. In the input analysis phase, the input features are converted into linguistic features such as phonemes; the acoustic model is responsible for converting linguistic features into acoustic features such as spectrogram. Then, the vocoder converts the spectrogram into audio signals. As technology has progressed, computer-based speech synthesis methods have evolved, starting with early methods such as articulatory synthesis \cite{coker1976model}, formant synthesis\cite{seeviour1976automatic}, and concatenative synthesis \cite{olive1977rule}, and moving on to more advanced methods such as statistical parametric speech synthesis (SPSS) \cite{yoshimura1999simultaneous} and neural network-based speech synthesis \cite{wang2017tacotron}. These neural network-based methods use deep learning techniques to generate speech and have shown great promise in producing highly realistic and natural-sounding speech. 

\cite{zen2013statistical} use SPSS to generate audio waveforms. This works by converting text into linguistic features and then using the DNN to generate acoustic features. This approach was better than the HMM-based speech synthesis model as it could model complex context dependencies. Instead of generating text using three different modules (analysis, acoustic and vocoder), Tacotron \cite{wang2017tacotron} is an end-to-end RNN-based, text-to-speech-system which uses an encode-decoder-based attention network and uses Griffin-Lim algorithm to generate the raw waveforms. Due to the recurrent nature of RNN, the encoder-decoder framework cannot be trained in parallel. With advancements in CNNs, authors proposed Deepvoice3 \cite{ping2018deep}, a fully convolutional neural network that generated mel-spectrogram instead of complex linguistic features. The growing use of transformers motivated the researchers to propose transformerTTS \cite{li2019neural}, which is based on an encoder-decoder-based attention network that can generate mel-spectrograms from phenomes. GANs are also used for speech synthesis. Multi-SpectroGAN \cite{lee2020multi} can be used to train the multi-speaker model with only the adversarial feedback. This model uses a conditional discriminator and can synthesize the mel-spectrogram without using re-construction loss between ground truth and the generated mel-spectrogram. The existing speech synthesis detection dataset, such as ASVSpoof2019 LA \cite{wang2020asvspoof} uses NN-based, SPSS-based TTS systems. The dataset uses WORLD \cite{kawahara2006straight} and STRAIGHT \cite{kawahara2006straight} as vocoders for generating the raw audio. The existing datasets are shown in Table \ref{tab:all_dataset}. However, the existing dataset lacks samples from sophisticated TTS systems that use diffusion models \cite{lee2021priorgrad}.

\subsection{Replay Attacks}
In a Replay attack, the attacker intercepts the speech of the legitimate user, records it and plays it back later to impersonate the original speaker. This is the easiest kind of attack as it does not require any technical expertise. The existing Replay attack generation techniques primarily focus on single-hop attacks. Datasets like ASVspoof 2017 \cite{kinnunen2017asvspoof} and ASVspoof 2019 \cite{wang2020asvspoof} only have single-playback recordings (i.e. played once), so they cannot assess anti-spoofing algorithms against multi-playback attacks. Also, the existing datasets need to consider microphone array characteristics, which are crucial in modern VCDs. Datasets like Voice Spoofing Detection Corpus (VSDC) \cite{baumann2021voice} handle this problem by proposing multi-hop replay attacks. The multi-hop replay attack problem can be considered a multi-class classification problem instead of the binary-class classification problem. This would assist in evaluating countermeasure systems in more realistic situations than in a lab environment. The existing Replay attack detection datasets are shown in Table \ref{tab:all_dataset}.

\subsection{Adversarial Attacks}

The adversarial attacks are performed by adding small and imperceptible perturbations to the original audio data that can fool the system into producing incorrect predictions. This started with \cite{szegedy2014intriguing} showing that adversarial examples can fool image classification models. This led to adversarial attacks becoming a topic of significant interest in image processing, leading to the development of numerous successful attack techniques, including the Fast Gradient Sign Method (FGSM) \cite{DBLP:journals/corr/GoodfellowSS14} and Basic Iterative Method (BIM) \cite{DBLP:conf/iclr/KurakinGB17a}.

Adversarial attacks in the audio field have gained growing attention from \cite{kreuk2018fooling}, following the progress made in adversarial attacks for images. There are two main categories of adversarial attacks on Speaker Recognition systems: optimization-based attacks \cite{DBLP:journals/corr/GoodfellowSS14,kreuk2018fooling} and signal processing-based \cite{DBLP:conf/interspeech/WangGX20} attacks. Optimization-based attacks generate adversarial examples by solving an optimization problem that formalizes the goal of the attack. This method typically involves minimizing a distance metric between the original and adversarial audio signals, subject to some constraints, such as the adversarial example being imperceptible to humans or having a small perturbation size. On the other hand, signal processing-based attacks use signal processing techniques to manipulate the original audio signal in a way that causes a desired misclassification by the SRS. These techniques can include adding noise, changing the frequency spectrum, or modifying the phase of the audio signal.

\section{Spoofed Speech Detection Techniques}
This section investigates different algorithms used to detect spoofing attacks in audio recordings. There are two main approaches to this problem, machine learning and deep learning-based algorithms. Most early spoof detection algorithms (before 2015) are based on traditional machine learning-based models (GMM, HMM, SVM). However, in recent years, deep learning models such as Convolutional Neural Networks (CNN) and Recurrent Neural Networks (RNN) have become increasingly popular for this task.

\subsection{Traditional ML-based Approaches}
\cite{satoh01_eurospeech} propose a synthetic speech detector to protect text-based ASV systems. The proposed model calculates the inter-frame difference of log-likelihood scores of the speaker's Gaussian Markov Model. The proposed architecture is only evaluated for the HMM-based speech synthesis model. This was the first-ever attempt to detect voice-based spoofing. This method was not evaluated for Replay attack detection. For replay attack detection, a similarity-based model \cite{wei_scorenormalize} is proposed, which calculates the similarity between test occurrence and recordings of the speaker. The main drawback of the paper is that it needs N-recorded speech for attack detection. \cite{alegre2013spoofing} use long-range features for synthetic speech detection. The authors extract utterance-level features extracted from frame-level features. The authors also show that certain ASV systems are more robust to spoofing attacks. The electromagnetic interference of audio recording creates an Electric Network Frequency (ENF) signal. \cite{enf_hui} propose a decorrelation operation to extract ENF signal, which can be used for Replay attack detection. 

Prior to 2015, most of the research on speech spoofing detection was conducted on private datasets, preventing the researchers from benchmarking the results. This led to the creation of ASVspoof2015 \cite{wu2015asvspoof}, which contains samples for TTS and VC attacks. \cite{todiscocqcc} extracted  Constant-Q Cepstral Coefficients (CQCCs) features based on constant Q transform (CQT) to detect spoofing attacks of the ASVSpoof2015 dataset. The authors have evaluated the proposed system on two existing datasets and the results have highlighted the sensitivity towards attacks. The authors did not evaluate their CQCC features on Replay attacks. Later, a new dataset ASVspoof2017 \cite{wu2017asvspoof} is prepared to spearhead the research in detecting Replay attacks.

\cite{font2017experimental} compared different handcrafted features such as Constant-Q Cepstral Coefficients (CQCCs) and Mel Frequency Cepstral Coefficients (MFCCs) to train a GMM-based model. The authors found that Subband Spectral Centroid Magnitude Coefficients (SCMCs) performed better than all other features. However, the model could not generalize when evaluated on the cross-corpus database. \cite{yang_octave} propose using power information of CQCC features using multi-level transform. The multi-level transform is applied to the power information of CQCC features.

In 2019, a more diverse dataset, ASVSpoof2019 \cite{wang2020asvspoof}, is released, containing synthetic attacks from 6 known and 13 unknown systems. The dataset also contains replayed attacks from 6 known systems. The researcher \cite{Malik2020ALight} showed that replay attacks could be modelled as a non-linear process, creating harmonic distortion in the signal. These harmonic distortions can be exploited to detect the attacks. The authors also propose a multi-hop replay attack dataset VSDC. The replayed attack samples of the ASVspoof2019PA dataset are replayed once and are easier to detect than the multi-hop replay attacks. A fusion of different features like MFCC, GTCC, Spectral Flux, and Spectral Centroid is explored for audio spoofing detection. The proposed architecture is only evaluated for synthetic speech detection.

\subsection{Deep Learning-based Approaches}
Deep learning-based techniques are extensively used for spoofing detection. The deep-learning-based approaches are used both for feature extraction and as classifiers. \cite{chen2015robust} have used deep learning for the audio spoofing detection task for the first time. This architecture uses deep learning architecture to extract learned representation. The authors compute a compact feature representation which uses distance metrics for classification. This algorithm extracts \textit{s-vector} features from Deep Neural Network (DNN) and uses Mahalanobis distance for spoofing detection. Convolutional long short-term neural network (CLDNN) \cite{endtoend2017} uses raw audio signal for spoof detection. The CLDNN uses layers that learn dependencies between time and frequency Domains. The architecture is evaluated on ASVspoof2015 and BTAS2016 datasets. However, the proposed architecture is used for both replay and speech synthesis attacks. Replay noise \cite{jin_noise} is explored by researchers for replay attack detection. The authors propose a Multi-Task learning framework to exploit this replay noise. However, the multi-task architecture is only evaluated for a single dataset. 

Residual connection-based approaches are also explored for audio spoofing detection. \cite{Alzantot} have built three different variants of residual networks. The authors observe that the fusion of models outperforms individual models. The fusion model achieves zero EER and zero t-DCF on the development set of the ASVspoof2019LA dataset. The zero EER and t-DCF show that the model is overfitting on the development set. The results show that fusion of five networks performed better than single systems and baseline architectures. This architecture can detect both replay and speech synthesis attacks. Authors have yet to experiment with the unified model. A deep neural network (DNN) based pipeline \cite{lai2019assert} is proposed for spoof detection. The proposed pipeline uses four stages feature engineering, DNN models, network optimization and system combination. The authors have explored multiple features and different variants of residual networks and squeeze-excitation networks. \cite{wu2020defense} showed that countermeasure system against adversarial PGD attack and proposed spatial smoothing and adversarial training \cite{wu2020defense} defence against an adversarial attack on countermeasure systems trained on ASVspoof2019 dataset. 

Several countermeasures use handcrafted features extracted from a raw audio signal to train deep neural networks.
Motivated by the vision transformers \cite{DBLP:conf/iclr/DosovitskiyB0WZ21}, \cite{DBLP:conf/ih/ZhangYZ21} use transformers and residual networks to detect fake speeches. The encoder of the transformer is used to extract deep features, which the residual network uses for classification. \cite{ranjan_ijcb} uses Spectral and Temporal features extracted from RawNet2 encoder for audio spoofing detection.

A capsule network with modified-routing algorithm \cite{DBLP:conf/icassp/LuoLLK021} is proposed for audio spoofing detection. The authors argue that capsules can learn better representations than convolutional networks. Motivated by residual connections, Several architectures \cite{Tak2} based on ResNet are proposed. 
The paper \cite{DBLP:conf/icassp/Tak0TNEL21} proposes several modifications to RawNet2 architecture for audio spoofing detection. \cite{jung2022aasist,Tak2} have also used Graph Neural Networks (GNNs) for spoofing tasks. \cite{jung2022aasist} propose a spectral and temporal graph attention layer for spoof detection. The architecture achieved state-of-the-art performance on the AVSpoof2019LA dataset. The authors have also proposed lightweight architecture for the deployment of edge devices. The architecture is not evaluated for Replay attacks.
Anti-spoofing systems only detect spoofing, but a combined system can be developed by incorporating speaker verification. Spoofing-aware speaker verification (SASV) challenge \cite{jung22c_interspeech} for the first time integrated speaker-verification and anti-spoofing. \cite{zhang2022probabilistic} propose a probabilistic fusion method for joint integration of ASV and CM systems. The authors use a product and fine-tuning strategy to get the SASV scores. Due to a limited dataset for evaluating the ASV and CM systems, the generalizability of the system could not be evaluated. Lightweight countermeasure \cite{liao22_spsc} is proposed for developing countermeasures systems for edge devices. Authors use Speaker distillation and generalized pre-training to train ResNetSE architecture effectively. With advancements in attack generation techniques, the future development of countermeasure systems lies in end-to-end deep 
learning-based countermeasure systems.

\subsection{Meta-Analysis}
\begin{figure}[t]
\centering
  \includegraphics[width=0.475\textwidth]{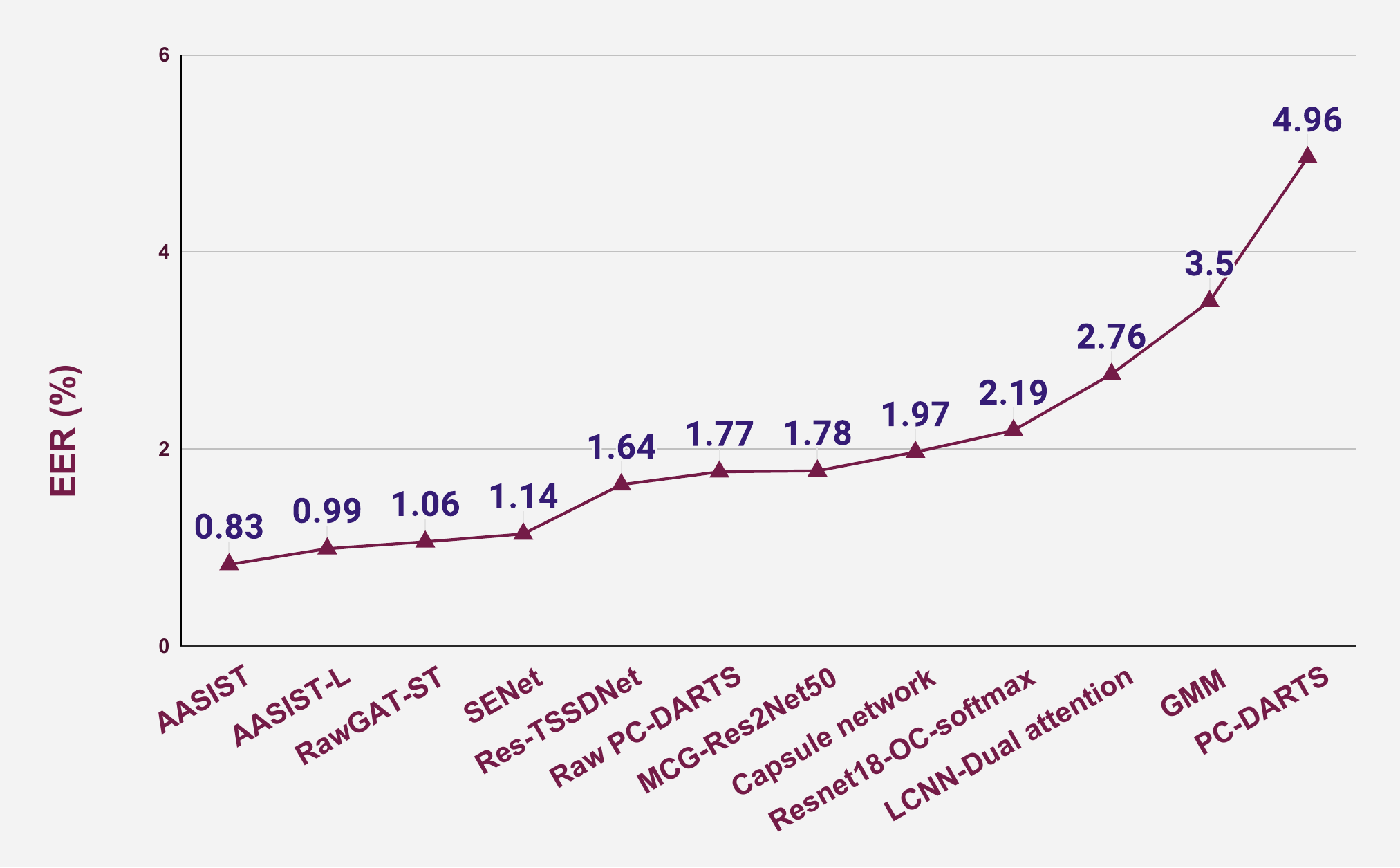}
  
  \caption{Performance of countermeasure system on the ASVSpoof2019LA dataset. The dataset contains samples of TTS and VC attacks.}
  \label{fig:sota_la}
  \includegraphics[width=0.475\textwidth]{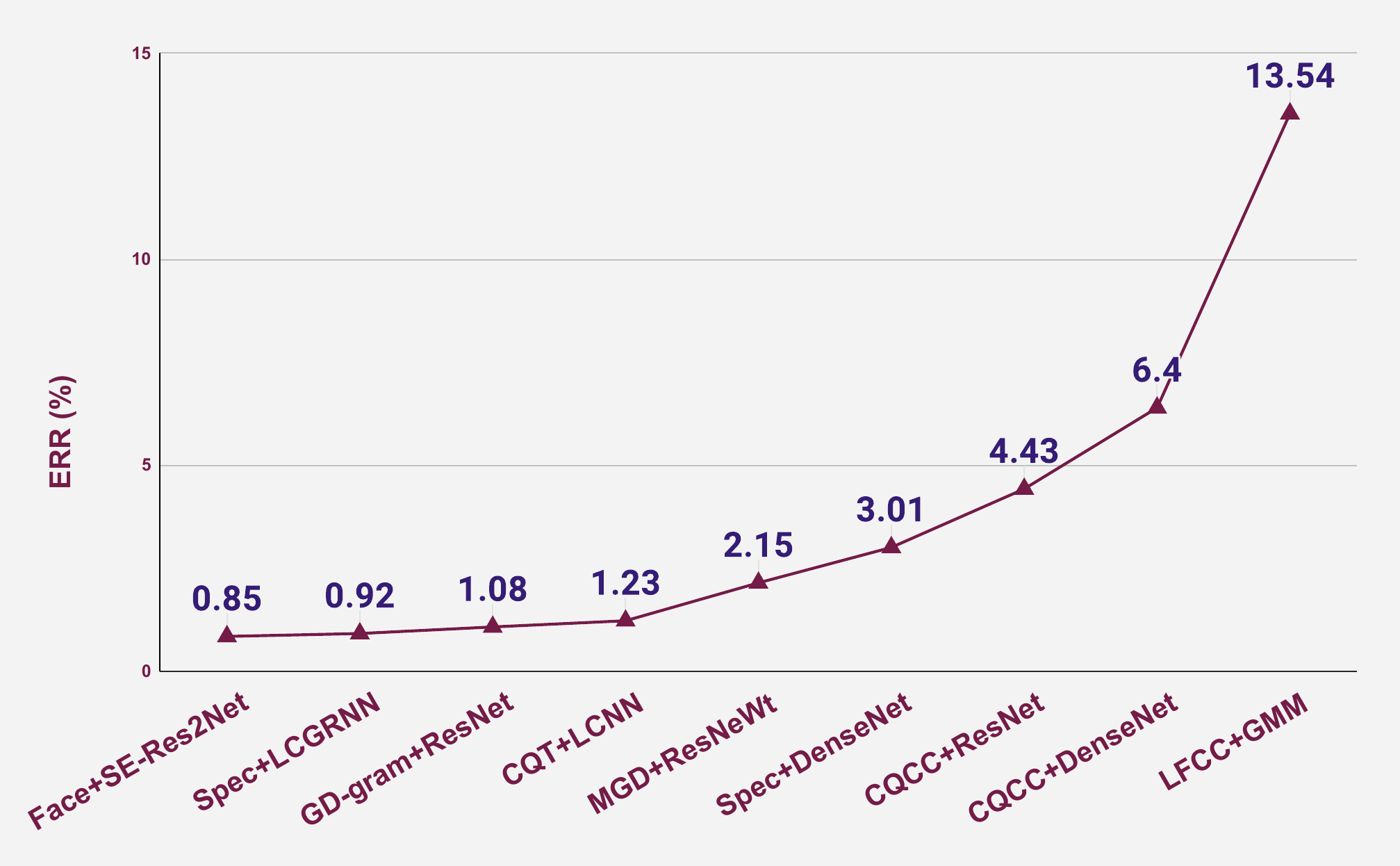}
  \caption{Performance of countermeasure system on the ASVSpoof2019PA dataset. The dataset contains samples of Replayed attacks.}
  \label{fig:sota_pa}
\end{figure}

In this section, we discuss the trends in developing countermeasure systems. We rank architectures with respect to their performance in detecting different attacks. 
\paragraph{Logical Access Attacks:} These included TTS and voice cloning attacks. The ASVSpoof2019LA dataset contains samples of logical access attacks. Figure \ref{fig:sota_la} shows the performance of different architectures on the ASvspoof2019LA dataset. Four of the five best systems use raw waveforms as input to the architecture, and the remaining one \cite{Zhang_silence} takes spectrogram as input. This shows that raw audio waveforms have complementary information with respect to handcrafted features. All the top five performing architectures used residual connections in their architecture. All top 3 performing architectures are based on raw audio and Graph neural networks. AASIST-L \cite{jung2022aasist} is a lightweight model with just 85K parameters and it is in the top 2 performing systems. CQCC features are the most informative features used by the authors in architecture MCG-Res2Net50 \cite{xu_channel}. However, none of the architecture is evaluated on a cross-corpus database. This limits the generalizability of the architectures. The results suggest that raw audio waveforms and deep learning-based systems can be used to build a countermeasure system for logical access attack detection.

\paragraph{Physical Access Attacks:} 
These generally contain samples of Replay attacks. The attack samples are collected in 27 different acoustic environments. Figure \ref{fig:sota_pa} shows the result for different architectures on the ASVSpoof2019PA dataset. The results suggest that hand-crafted features are essential for audio spoofing detection. The features extracted using CQT are extensively used for spoofing detection. Many architectures use ResNet and Light CNN as backbone classifiers. However, most systems are not evaluated on multiple attacks and cross-database settings.

\section{Open Challenges}

This section discusses the challenges and future prospects for research in ASV systems.
\subsection{Diverse Dataset Collection} Several challenges are associated with collecting datasets that can be used to train the countermeasure system for spoof detection. The challenges arise while collecting real audio data and generating fake audio. We list the key challenges as follows: 

\noindent\textit{Diversity of Dataset:} To effectively detect spoofing attacks, it is crucial to have a diverse set of audio data that accurately reflects the target population. In the case of India, this would mean taking into account a wide range of demographic factors such as gender, age, dialect, and language. It is essential to have a balanced representation of different genders and age groups in the dataset, to ensure that the detection algorithms can accurately identify spoofing attacks regardless of the speaker's gender or age. Dialects and language also play a crucial role in Indian society, with a vast array of regional dialects and multiple official languages. Hence, a balanced representation of the population is also essential. 

\noindent\textit{Quality of Data:} The quality of the data and the labelling of that data are critical to the success of an audio spoof detection task. A high-quality dataset should have accurate and detailed annotations, including speaker identity, gender, age, and dialect. Additionally, the recording conditions should be standardized, with high-quality equipment and real-world environments. A model only trained on data collected in a controlled environment may not perform well in real-world scenarios due to differences in acoustic conditions, background noise, and other environmental factors. 

\noindent\textit{Privacy Preservation:} Privacy concerns are a significant challenge in collecting bonafide audio data for audio spoof detection. Audio data can contain personal information that needs to be protected, such as names, addresses, and sensitive information. Data collection must comply with legal constraints for data protection, such as the European General Data Protection Regulation (GDPR). Privacy regulations aim to protect personal information from unauthorized access and misuse, and researchers must ensure that the data collection process adheres to these regulations. 

\noindent\textit{Accent-Aware Synthetic Data:} Synthetic data should reflect the language and accent of the target speaker to be impersonated, making cross-lingual compatibility an essential factor to consider in creating synthetic data. The synthetic data should also reflect those specific accents if the target speaker speaks English with a different accent, such as an Indian or Chinese accent. This requires a comprehensive understanding of different accents and their characteristics and using language-specific tools and techniques to create synthetic data that accurately reflects each accent. If the synthetic data does not reflect the target speaker's accent, it may lead to incorrect results in the detection algorithms or biases in the performance of the algorithms. 

\noindent\textit{Multi-Speaker Dataset:} One of the biggest challenges while collecting synthetic data for low-resource languages like Hindi is the limited availability of high-quality speech data. In many cases, the amount of data available for low-resource languages is significantly smaller than that for high-resource languages, which can lead to overfitting and poor generalization performance. So, to perform better, synthetic data should reflect the language and accent of the target speaker to be impersonated, making cross-lingual compatibility an essential factor to consider in creating synthetic data. 

\subsection{Robust Spoof Detection Algorithms}
Most of the existing detection algorithms need to be more generalizable. Spoof detection models are evaluated on limited and non-diverse data, leading to an inaccurate assessment of their generalization ability. Existing spoof detection models outperform benchmark datasets but fail in real-world scenarios. Also, the existing evaluation metrics must adequately capture the models' generalization ability.
\subsection{Cross-Corpus Evaluation}
Most existing spoof detection models are not evaluated on cross-corpus, which is a crucial step in developing audio spoof detection models. It helps to determine their generalizability and robustness to variations in recording conditions, data distribution, and noise types. Such evaluations are necessary because a model trained on a specific corpus may perform well on that particular dataset but not on others due to different data distribution and recording conditions. \cite{korshunov16_interspeech} show that the existing spoofing detection systems need to be more generalizable as they perform poorly when tested on the same attack samples from a different dataset. The data variability can lead to significant differences in model performance across different corpora, making it essential to evaluate models against diverse datasets.

\subsection{Language Independent Models} People use speaker verification systems in their native languages to authenticate themselves. However, existing countermeasure systems are heavily language dependent. This poses a severe risk to the ASV systems. This possible solution is that separate models must be trained and deployed for each language, which can be time-consuming and computationally intensive, increasing the complexity of building and deploying audio spoof detection systems. Developing language-independent models that can effectively detect audio spoofing across multiple languages would greatly benefit the field and increase the efficiency of audio spoof detection systems.
\subsection{Universal Spoof Detection Model} There exist a few models that can detect all types of spoofing attacks (replay, TTS, VC, adversarial attacks). This is because different types of attacks require different methods of detection. For example, detecting replay attacks might require examining the properties of a microphone, whereas detecting adversarial examples might require identifying changes in the statistical properties of the data. Therefore, creating a universal model for all types of attacks is difficult. The universal model can detect a broader range of attacks, improving security in the field and making attacks less likely to go unnoticed.
Additionally, deploying a single model costs less than deploying multiple models. This is because it reduces the need for multiple systems and personnel to manage and maintain each model. However, a general-purpose model may not perform as well as a specialized model for a particular type of attack, as it requires a trade-off between accuracy and performance. 
\subsection{Privacy Preserving Spoof Detection Models}
As data privacy becomes increasingly essential, Federated Learning (FL) provides a solution that allows organizations to collaborate and share information while preserving the privacy of individual data. This is particularly relevant in tasks where biometric information, such as speech, is collected. FL enables multiple parties to collaborate to train a neural network without exchanging data samples, ensuring individual data privacy. However, there are various challenges while training a federated learning-based spoof detection model. One of the significant challenges is non-IID data distribution, where each participant has their own data distribution, which may differ from other participants' data distribution. This results in a non-IID data scenario and can lead to less effective learning compared to centralized training methods.

\subsection{Interpretable and Explainable Models}
Explainability is very important in developing spoofed detection models, as it allows us to better understand these models' decision-making processes. This is important for several reasons. First, explainability helps ensure that models are not biased toward specific classes or features that can negatively impact performance. It then provides insight into how the model processes the input data so that further improvements can be made. Finally, it gives a better understanding of model limitations, and helps identify and fix potential errors. For example, in the image domain, CNNs have given excellent results for tasks such as image classification. However, the decision-making process of these models still needs to be improved to interpret. To address this, researchers proposed using explainable AI (XAI) techniques such as Grad-CAM (Gradient-weighted Class Activation Mapping)\cite{DBLP:Journals/ijcv/SelvarajuCDVPB20} which provides a visual explanation of CNN prediction. This gives a better understanding of the features that the model uses to make predictions. Along the same lines, \cite{ge2022explainable} use SHapley Additive exPlanations (SHAP) to describe a detection model for audio spoofing. However, this method only works for confirmed attacks. 

\subsection{Fair Spoof Detection}
Fairness and bias are crucial in developing spoof detection models as they directly affect the performance and accuracy of the models. Bias in these models can result in unequal treatment of different groups, leading to unfair and unreliable decisions. To ensure that these models are fair, it is essential to consider demographic information and various factors that may impact the accuracy of the model's predictions. Techniques such as debiasing the training data and using gender-neutral and diverse datasets can be employed to address this. However, there is no dataset to evaluate the fairness of the existing countermeasure system. The research community needs a dataset that has labels about the gender, ethnicity, and accent of the speakers.

\section{Conclusion}
The advancements in attack generation methods have motivated researchers to develop countermeasures systems for audio spoofing detection. In this paper, we review the existing literature for audio spoofing detection. Despite the advancements in audio spoofing detection methods, several research gaps exist. We discuss the research gaps and open challenges that require proper attention and continuous research efforts. Solving or providing solutions for these open research problems is necessary for building trusted audio spoofing detection systems.

\section*{Acknowledgements}
This research is supported by a grant from MHA, India. M. Vatsa is also partially supported through the SwarnaJayanti Fellowship by the Government of India.

\bibliographystyle{named}

\end{document}